\documentclass[letterpaper]{article} 
\usepackage{aaai2026}  
\usepackage{times}  
\usepackage{helvet}  
\usepackage{courier}  
\usepackage[hyphens]{url}  
\usepackage{graphicx} 
\usepackage{natbib}  
\usepackage{caption} 
\usepackage{algorithm}
\usepackage{algorithmic}
\usepackage{amsmath}
\usepackage{booktabs}
\usepackage{multirow} 
\usepackage{makecell}

\usepackage{newfloat}
\usepackage{listings}
\DeclareCaptionStyle{ruled}{labelfont=normalfont,labelsep=colon,strut=off} 
\floatstyle{ruled}
\newfloat{listing}{tb}{lst}{}
\floatname{listing}{Listing}
\usepackage{tcolorbox}

\newtcolorbox{myprompt_single}[2][]
{
    colframe=black,      
    colback=gray!20,     
    boxrule=1pt,         
    arc=4pt,             
    left=10pt,           
    right=10pt,
    top=10pt,            
    bottom=10pt, 
    title=#2
}

\newtcolorbox{myprompt_double}[2][]
{
    colframe=black,      
    colback=gray!20,     
    boxrule=1pt,         
    arc=4pt,             
    left=10pt,           
    right=10pt,
    top=10pt,            
    bottom=10pt, 
    width=\textwidth,
    title=#2
}

\title{Cog-RAG: Cognitive-Inspired Dual-Hypergraph with Theme Alignment Retrieval-Augmented Generation}
\author{
    Hao Hu\textsuperscript{\rm 1}, 
    Yifan Feng\textsuperscript{\rm 2},
    Ruoxue Li\textsuperscript{\rm 3},
    Rundong Xue\textsuperscript{\rm 1},
    Xingliang Hou\textsuperscript{\rm 4},\\ 
    Zhiqiang Tian\textsuperscript{\rm 4}, 
    Yue Gao\textsuperscript{\rm 2},
    Shaoyi Du\textsuperscript{\rm 1}\thanks{Corresponding Authors.},
}

\affiliations{
    \textsuperscript{\rm 1}State Key Laboratory of Human-Machine Hybrid Augmented Intelligence, National Engineering Research Center for Visual Information and Applications, and Institute of Artificial Intelligence and Robotics, Xi'an Jiaotong University\\
    \textsuperscript{\rm 2}BNRist, THUIBCS, BLBCI, School of Software, Tsinghua University\\    
    \textsuperscript{\rm 3}School of Artificial Intelligence, Xidian University\\ 
    \textsuperscript{\rm 4}School of Software Engineering, Xi'an Jiaotong University\\
    {huhao@stu.xjtu.edu.cn}, {dushaoyi@xjtu.edu.cn}

}

\usepackage{bibentry}

\begin{document}

\maketitle


\begin{abstract}
Retrieval-Augmented Generation (RAG) enhances the response quality and domain-specific performance of large language models (LLMs) by incorporating external knowledge to combat hallucinations. In recent research, graph structures have been integrated into RAG to enhance the capture of semantic relations between entities. However, it primarily focuses on low-order pairwise entity relations, limiting the high-order associations among multiple entities. Hypergraph-enhanced approaches address this limitation by modeling multi-entity interactions via hyperedges, but they are typically constrained to inter-chunk entity-level representations, overlooking the global thematic organization and alignment across chunks. Drawing inspiration from the top-down cognitive process of human reasoning, we propose a theme-aligned dual-hypergraph RAG framework (Cog-RAG) that uses a theme hypergraph to capture inter-chunk thematic structure and an entity hypergraph to model high-order semantic relations. Furthermore, we design a cognitive-inspired two-stage retrieval strategy that first activates query-relevant thematic content from the theme hypergraph, and then guides fine-grained recall and diffusion in the entity hypergraph, achieving semantic alignment and consistent generation from global themes to local details. Our extensive experiments demonstrate that Cog-RAG significantly outperforms existing state-of-the-art baseline approaches. 
\end{abstract}

\section{Introduction}
Retrieval-Augmented Generation (RAG) has recently gained increasing attention for enhancing the performance of large language models (LLMs) on knowledge-intensive tasks \cite{lewis2020retrieval, gao2023retrieval, li2024structrag}. It combats LLMs' hallucination by incorporating external knowledge, thereby enhancing response quality and reliability \cite{ayala2024reducing, xia2025improving}. Moreover, it enables integration with private or domain-specific knowledge bases, thereby increasing the model’s adaptability to vertical domains. With these advantages, RAG has emerged as a fundamental component in question answering, document understanding, and intelligent assistants \cite{fan2024survey, dong2025toward}.

Despite the notable potential of RAG in enhancing LLMs' response quality, current methods mostly rely on a flattened chunk-based retrieval that matches queries to document chunks via vector similarity \cite{asai2023self, yang2024crag}. However, this fails to capture inter-chunk dependencies and semantic hierarchies, resulting in fragmented and weakly connected retrieval content, which weakens the model’s structured understanding of the entire knowledge.

\begin{figure}[t]
    \centering\includegraphics[width = \linewidth]{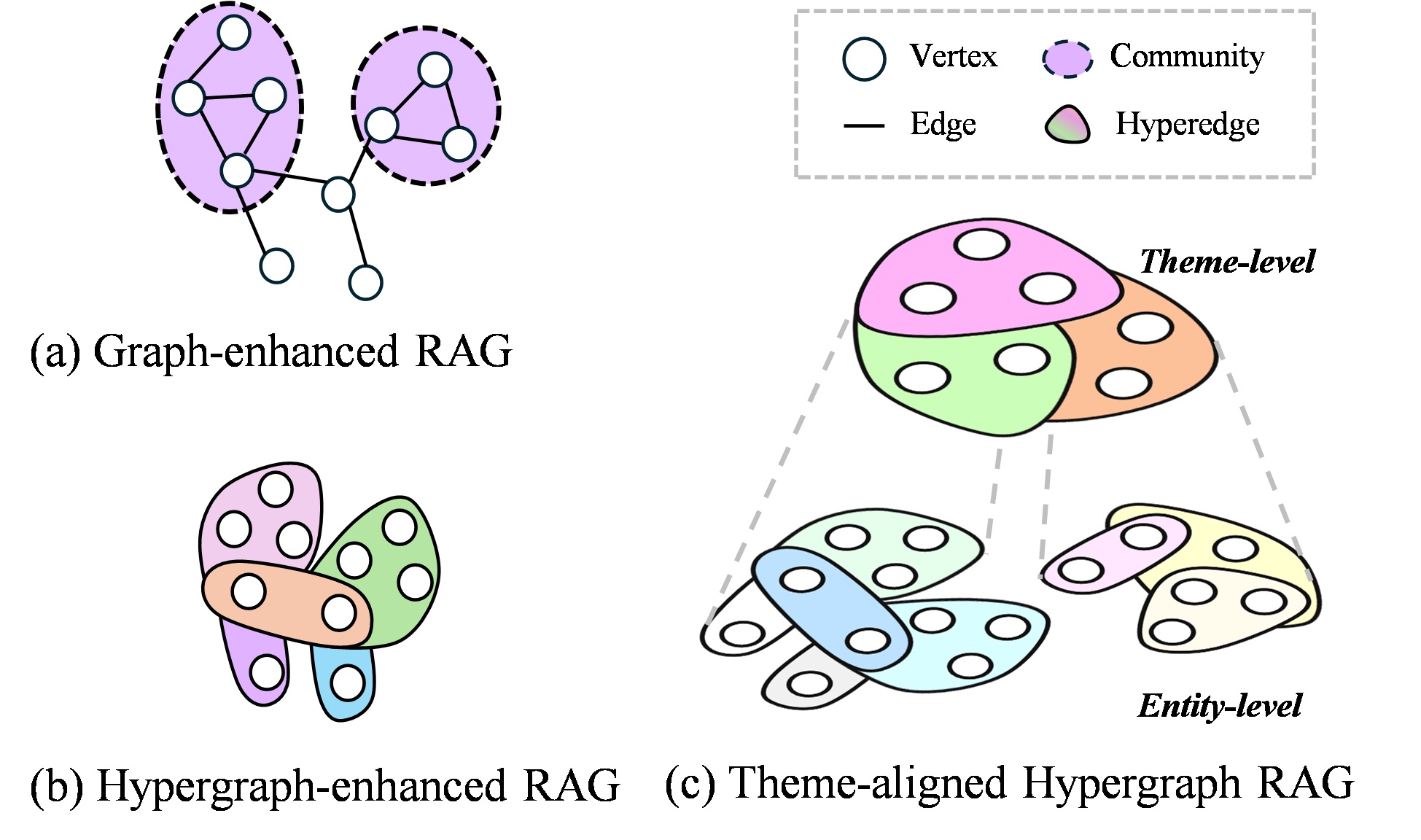}
    \caption{Knowledge modeling of graph, hypergraph, and our theme-enhanced RAG.}
    \label{fig1}
\end{figure}

To address this, recent studies have attempted to introduce graph structures into RAG framework, aiming to construct corpus-wide knowledge graphs that capture the structural semantic relations between entities \cite{peng2024graph, zhang2025survey, wang2025archrag}. For instance, GraphRAG \cite{edge2024local} and LightRAG \cite{guo2024lightrag} utilize graph structures to strengthen entity-level indexing and retrieval, explicitly capturing semantic relations to improve information organization. Hyper-RAG \cite{feng2025hyper} uses hypergraphs to model complex relations between multiple entities. Nevertheless, these approaches primarily concentrate on entity-level structural modeling and lack a unified organization of knowledge themes and semantic-driven reasoning, making it difficult to support hierarchical integration of information from macro comprehension to micro details.

It is worth noting that humans tend to follow a top-down information processing path when handling complex tasks \cite{cheng2025human, gutiérrez2024hipporag}. They begin by identifying the core themes of the problem and constructing a global semantic scaffold. Based on this, they recall and integrate relevant details to form a coherent and structured response. 
This ``theme-driven, detail-recall" cognitive pattern reflects the inherent hierarchical organization and semantic coherence in human information processing.

Inspired by this cognitive insight, we propose a dual-hypergraph with theme alignment RAG framework (Cog-RAG). Figure \ref{fig1} shows its difference with other methods in knowledge modeling. Our method leverages a dual-hypergraph structure to model the global theme structure and fine-grained high-order semantic relations. In addition, it introduces a cognitive-inspired two-stage retrieval strategy that simulates the human top-down information comprehension process, thereby enhancing the semantic consistency and structural expressiveness of generated responses. The main contributions are summarized as follows:
\begin{itemize}
\item We propose Cog-RAG
that simulates the human top-down information processing path, enabling hierarchical generation modeling from macro-level semantic comprehension to micro-level information integration.
\item We design a dual-hypergraph semantic indexing scheme to separately model global inter-chunk theme structure and intra-chunk fine-grained high-order semantic relations, overcoming the limitations of prior graph-enhanced RAG models that focus only on pairwise relations and lack unified thematic organization.
\item We develop a cognitive-inspired two-stage retrieval strategy that first activates relevant context in the theme hypergraph and then triggers detail recall and diffusion in the entity hypergraph. This ``theme-driven, detail-recall" process enables semantic alignment across granularity and significantly improves the coherence and quality of the response.
\end{itemize}

\section{Related Work}
\subsection{RAG with Knowledge Graph}
Most text-based RAG methods \cite{asai2023self, zhang2024arl2, xia2025improving,yang2024crag} rely on a flattened paragraph structure, which makes it difficult to model semantic associations and contextual dependencies across text chunks, thereby limiting the accuracy and completeness of generated responses. 
To address this issue, recent studies \cite{sarmah2024hybridrag, peng2024graph} have explored knowledge graphs within the RAG framework to structurally represent entities and relations, aiming to enhance the organization and semantic expressiveness of retrieved content. 

Some recent studies \cite{gutiérrez2024hipporag,li2023leveraging, cheng2025human} attempt to automatically extract knowledge graph triples from the corpus and retrieve relevant subgraphs to improve content relevance and interpretability. However, these methods typically construct sparse graph structures, making it difficult to capture the full semantic space and contextual dependencies. 
To address the semantic sparsity issue and better model the semantic structure of documents, graph-enhanced RAG approaches \cite{edge2024local, guo2024lightrag, chen2025pathrag} extract entities and their relations, and directly build document-level graph databases enriched with contextual information, thereby reducing information loss during the text-to-graph conversion process. GraphRAG and LightRAG employ LLMs to extract entities and relations from texts as vertices and edges of the graph. 
Nevertheless, existing methods primarily focus on low-order pairwise relations between entities, neglecting high-order group associations and global topic modeling, which limits the semantic coverage and structural expressiveness of the generated content.

\subsection{Hypergraph}
Hypergraphs connect multiple vertices via hyperedges, effectively modeling complex high-order relationships among entities and overcoming the limitation of conventional graphs, which support only binary relations \cite{gao2022hgnn+, feng2024beyond}. These strong modeling capabilities have led to significant progress in fields such as recommender systems, social network analysis, and brain network modeling \cite{ji2020dual, sun2023self, han2025hypergraph}.
However, in the RAG framework, existing research is constrained to graph structures, primarily focusing on the pairwise relationships between entities. To model multiple entity group semantic associations, GraphRAG generates community reports through the semantic clustering of entities, while HiRAG \cite{huang2025retrieval} incorporates hierarchical graph knowledge via multi-level clustering. While effective in capturing local relationships, these methods rely on discrete category divisions and fail to model higher-order dependencies, resulting in information loss.

In contrast, hypergraphs naturally connect multiple entities through hyperedges, allowing them to interact with multiple hyperedges at once. This enables the capture of higher-order dependencies in a unified framework. It avoids the fragmentation and loss of information typical in clustering approaches and maximizes the retention of semantic information during text-to-graph conversion \cite{feng2025hyper,luo2025hypergraphrag}. The hypergraph structure enhances semantic associations both within and across documents, thereby improving the RAG system’s ability to understand context and ensure consistency in generated responses.

\begin{figure*}[t]
    \centering\includegraphics[width = 0.95\linewidth]{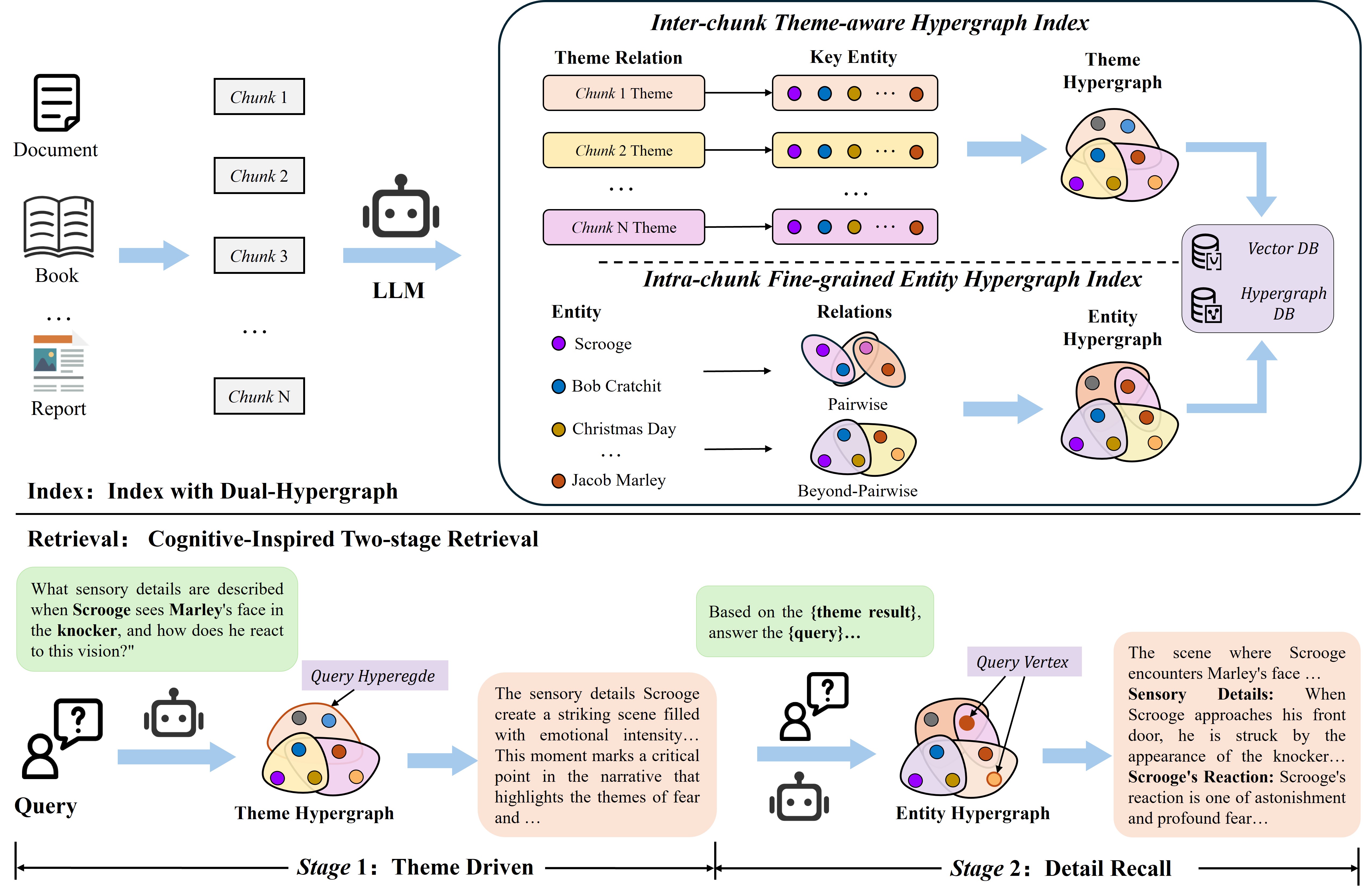}
    \caption{The overall framework of Cog-RAG.}
    \label{fig2}
\end{figure*}

\section{Preliminary}
In this section, we provide a general expression for RAG and graph-enhanced RAG, referring to the definitions in \cite{edge2024local,guo2024lightrag}.

An RAG system $\mathcal{M}$ generally includes LLM, retriever, and corpora, which can be defined as follows:
\begin{equation}
    \mathcal{M} = (LLM,\mathcal{R}(q,\mathcal{D})).
\end{equation}

Given a query $q$, the retriever $\mathcal{R}$ selects relevant contexts from the corpora $\mathcal{D}$, which are then used by the LLM to generate a response.

For the graph-enhanced RAG, the corpus is organized into a graph structure, where vertices represent entities and edges represent the relations. It can be formally defined as follows:
\begin{equation}
    \mathcal{M} = (LLM,\mathcal{R}(q,\mathcal{D}=\{\mathcal{V},\mathcal{E}\})).
\end{equation}

The query $q$ retrieves relevant vertices or edges from the graph-structured corpus $\mathcal{D}=\{\mathcal{V},\mathcal{E}\}$, enabling the LLM to respond.

\section{Method}
\subsection{Overview}
As illustrated in Figure \ref{fig2}, Cog-RAG comprises two main components: dual-hypergraph indexing and cognitive-inspired two-stage retrieval. We construct the dual-hypergraph with complementary semantic granularity: the theme hypergraph captures semantic theme associations between chunks (such as storyline, narrative outline, and summary), providing global semantic theme organization; the entity hypergraph models fine-grained high-order relations among entities (such as persons, concepts, and events), supporting local semantic relations.
In the retrieval stage, mimicking the human “top-down” reasoning pattern, Cog-RAG first activates relevant themes in the theme hypergraph as global semantic anchors. Guided by these anchors, it then retrieves related entities and relations information from the entity hypergraph. The final response is generated via LLMs, utilizing theme-driven, detail-recall knowledge as evidence.

\subsection{Dual-Hypergraph Indexing}
To more effectively model complex high-order associations among multiple entities in corpora and avoid the information loss by graph structure, we introduce hypergraphs for modeling. The general formulation is defined as follows:
\begin{equation}
    \mathcal{M} = (LLM,\mathcal{R}(q,\mathcal{D}=\{\mathcal{V},\mathcal{E}_\text{low},\mathcal{E}_\text{high}\})),
\end{equation}
where hyperedges are used to represent relations. $\mathcal{E}_\text{low}$ denotes low-order pairwise entity relations, while $\mathcal{E}_\text{high}$ refers to high-order beyond pairwise multiple entities associations.

\subsubsection{Theme-Aware Hypergraph Index}
The theme hypergraph is designed to model the semantic storyline structure of a document, establishing a narrative outline that provides cognitive guidance for subsequent detail retrieval.

Given a corpus $\mathcal{D}$, such as books, reports, or manuals, we first segment it into a set of chunks using a fixed-length sliding window with partial overlap to maintain semantic integrity, denoted as:
\begin{equation}
    \mathcal{D}=\{D_1,D_2,...,D_N\},
\end{equation}
where $D_i$ denotes the $i$-th document chunk, serving as the basic unit for subsequent analysis.

Then, we perform semantic parsing on each chunk using LLMs to automatically extract its latent theme and associated key entities, thereby constructing a theme hypergraph.
Specifically, we first employ predefined theme-level extraction prompts $\mathcal{P_\text{ext\_theme}}$, $\mathcal{P_\text{ext\_key}}$ (detailed in Appendix) to guide the LLM in performing semantic parsing for each chunk $D_i$ and outputting the corresponding theme. Then, further extract the key entities related to the theme. The calculation process is as follows:
\begin{equation}
\left\{
\begin{aligned}
    &\mathcal{E_\text{theme}} = LLM(\mathcal{P_\text{ext\_theme}}(D_i)) \\
    &\mathcal{V_\text{key}} = LLM(\mathcal{P_\text{ext\_key}}(D_i,\mathcal{E_\text{theme}}))
\end{aligned}
\right.
\quad \text{for} \quad D_i \in \mathcal{D}.
\end{equation} 
Based on the extracted themes and entities, we can construct the theme hypergraph $\mathcal{G_\text{theme}}$, denoted as:
\begin{equation}
    \mathcal{G_\text{theme}} = \{\mathcal{V_\text{key}}, \mathcal{E_\text{theme}}\},
\end{equation} 
where each hyperedge $\mathcal{E_\text{theme}}$ represents the narrative theme of the chunk, while the vertices $\mathcal{V_\text{key}}$ are the key entities.

\subsubsection{Fine-Grained Entity Hypergraph Index}
After constructing the theme hypergraph, we obtain a global thematic structure among chunks. To further capture fine-grained multi-entity relations, we construct an entity hypergraph within each chunk to model high-order relations among entities, supporting subsequent fine-grained retrieval.

For each chunk $D_i$, we first extract entities (such as person, event, organization, etc.) and their descriptions using LLMs, which serve as the vertex set for the fine-grained entity hypergraph. Based on the semantic relations among these entities, we then construct two types of hyperedges: low-order hyperedges $\mathcal{E}_\text{low}$ capture basic pairwise relations, while high-order hyperedges $\mathcal{E}_\text{high}$ model more complex semantic associations among multiple entities, such as co-occurrence in events or causal links. The extraction process is represented as follows:
\begin{equation}
\left\{
\begin{aligned}
    &\mathcal{V} = LLM(\mathcal{P_\text{ext\_entity}}(D_i)) \\
    &\mathcal{E}_\text{low} = LLM(\mathcal{P_\text{ext\_low}}(D_i,\mathcal{V} ))\\
    &\mathcal{E}_\text{high} = LLM(\mathcal{P_\text{ext\_high}}(D_i,\mathcal{V})) 
\end{aligned}
\right.  
\quad \text{for} \quad D_i \in \mathcal{D},
\end{equation}
where $\mathcal{P_\text{ext\_entity}}$ refers to the prompt designed for entity extraction from the text. $\mathcal{P_\text{ext\_low}}$ and $\mathcal{P_\text{ext\_high}}$ (detailed in Appendix) represent the extraction of paired and group relations from the obtained entities, respectively. 

Finally, all extracted entities, along with their low-order and high-order relations, are organized into a fine-grained entity hypergraph $\mathcal{G_\text{entity}}$ and stored in a hypergraph database.
\begin{equation}
\mathcal{G_\text{entity}} = \{\mathcal{V},\mathcal{E}_\text{low},\mathcal{E}_\text{high}\}.
\end{equation}

\subsection{Cognitive-Inspired Two-Stage Retrieval}
Motivated by the top-down information processing pattern observed in human memory retrieval, 
we design a cognitive-inspired two-stage retrieval strategy.
Specifically, it first identify theme threads in the theme hypergraph related to the query. These threads then serve as cues to guide the retrieval of fine-grained information from the entity hypergraph.

For a given user query $q$, we first extract theme keywords (overarching concepts or themes), as follows:
\begin{equation}
\mathcal{X_\text{theme}} = LLM(\mathcal{P_\text{keyword}}(q)),
\end{equation}
where $\mathcal{X_\ast}=\{{x_1,x_2,...}\}$, $\mathcal{P_\text{keyword}}$ is the prompt for extracting theme keywords from the query, detailed in Appendix. 

\subsubsection{Theme-Aware Hypergraph Retrieval}
Subsequently, we perform structured retrieval over the hypergraph database. It is worth noting that theme keywords reflect abstract semantic relations among multiple entities and are therefore used to retrieve relevant hyperedges. 

Therefore, in the first stage of retrieval, the extracted theme keywords are used to perform semantic matching within the theme hypergraph and select the top-k relevant theme hyperedges.
\begin{equation}
\mathcal{E}_\text{rel} = \{\mathcal{R}(x_i, \mathcal{E}_\text{theme}) | x_i \in \mathcal{X}_\text{theme} \} ,
\end{equation}
where $\mathcal{E}_\text{rel}$ represents the relevant hyperedges retrieved from the vector database. Then, we perform a diffusion process over the hypergraph database to retrieve their neighboring vertices, providing additional context awareness for the retrieved theme.
\begin{equation}
    \mathcal{V}_\text{dif} = \{\mathcal{N}(e_i, \mathcal{G}_\text{theme}) | e_i \in \mathcal{E}_\text{rel} \} ,
\end{equation}
where $\mathcal{N}$ denotes the function of obtaining the corresponding neighbors from the hypergraph. $\mathcal{V}_\text{dif}$ is the diffusion vertices. Then, both the $\mathcal{E}_\text{rel}$ and $\mathcal{V}_\text{dif}$, along with the corresponding textual contexts, are fed into LLMs as prior knowledge to generate an initial theme-aware answer as follows:
\begin{equation}
    \mathcal{A}_\text{theme} = LLM(q,\mathcal{E}_\text{rel},\mathcal{V}_\text{dif},\mathcal{C}_\text{e\_rel},\mathcal{C}_\text{v\_dif}),
\end{equation}
where $\mathcal{A}_\text{theme}$ denotes the output of query $q$ after retrieving from $\mathcal{G}_\text{theme}$, $\mathcal{C_\ast}$ is the corresponding context.

\begin{figure*}[!ht]
    \centering\includegraphics[width = 0.96\linewidth]{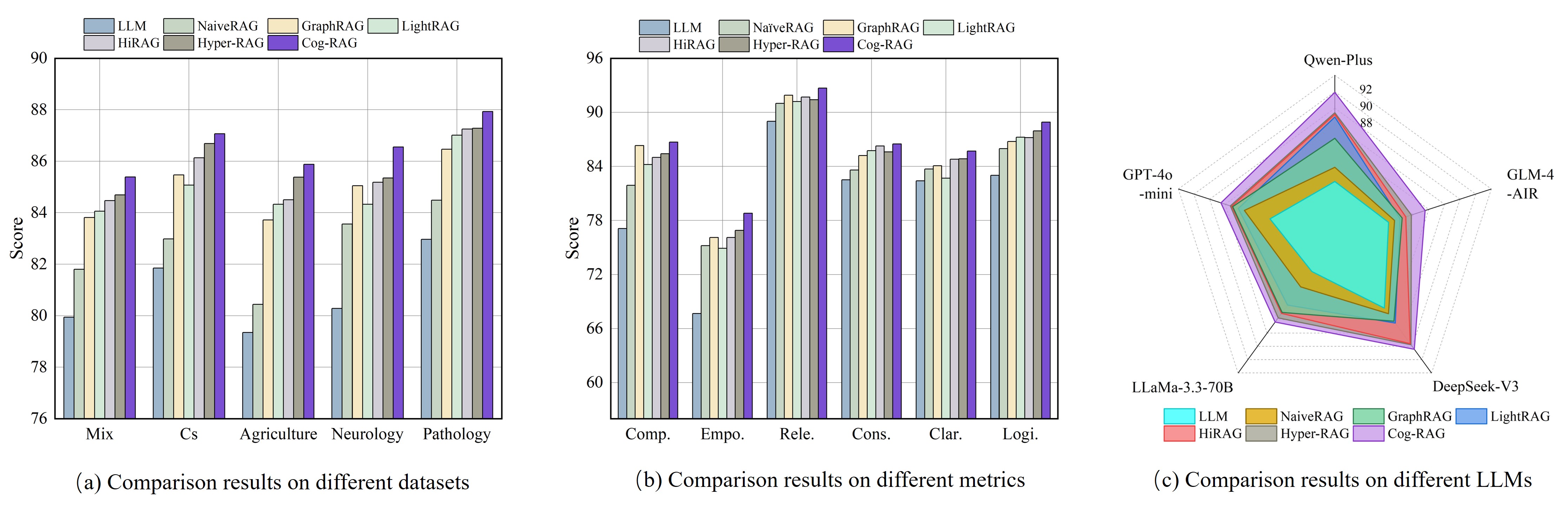}
    \caption{Test results by scoring. (a) is the comparison results on five datasets; (b) is the results of the neurology dataset on six dimensions; (c) shows the evaluation results on different LLMs.}
    \label{fig3}
\end{figure*}

\subsubsection{Theme-aligned Entity Hypergraph Retrieval}
After completing the initial theme-based retrieval, we further perform fine-grained information retrieval within the entity hypergraph. Guided by the retrieved themes, this section supplements entity-level semantic details, enabling effective alignment between local information and global themes.

Based on the theme response, we further extract the aligned entity keywords (specific entities or details) from query $q$, which is as follows:
\begin{equation}
\mathcal{X_\text{entity}} = LLM(\mathcal{P_\text{align}}(q,\mathcal{A}_\text{theme})),
\end{equation}
where $\mathcal{P}_\text{align}$ is the prompt (detailed in Appendix) used for extracting entity keywords aligned with the theme. $\mathcal{X_\text{entity}}$ primarily describe concrete individual information and are thus matched to vertices. The natural combination of two types of keywords and hypergraph structure enhances both retrieval specificity and structural compatibility.

Unlike the theme retrieval stage which targets hyperedges, this stage focuses on retrieving top-k vertices within the entity hypergraph by entity keywords, thereby achieving fine-grained semantic supplement and structured alignment. 
\begin{equation}
\mathcal{V}_\text{rel} = \{\mathcal{R}((x_i, \mathcal{V}_\text{entity}) | x_i \in \mathcal{X}_\text{entity} \} ,
\end{equation}
where $\mathcal{V}_\text{rel}$ refers the retrieved relevant entities. Then perform a hypergraph structure diffusion as follows:
\begin{equation}
\mathcal{E}_\text{dif} = \{\mathcal{N}(v_i, \mathcal{G}_\text{entity}) | v_i \in \mathcal{V}_\text{rel} \} .
\end{equation}
Finally, the retrieved $\mathcal{V}_\text{rel}$, diffusion $\mathcal{E}_\text{dif}$, and their corresponding contexts, integrated with the previous theme information $\mathcal{A}_\text{theme}$, to form a structured input for LLMs to generate the final answer $\mathcal{A}$ for query $q$, thereby achieving a comprehensive semantic generation process from theme guidance to detailed support.
\begin{equation}
    \mathcal{A}_\text{} = LLM(q,\mathcal{A}_\text{theme}, \mathcal{V}_\text{rel},\mathcal{E}_\text{dif},\mathcal{C}_\text{v\_rel},\mathcal{C}_\text{e\_dif}).
\end{equation}

\section{Experiments}
\subsection{Experimental Setup}
\subsubsection{Datasets}
To systematically evaluate our method across diverse application scenarios, we adopt five datasets from two benchmarks: Mix, CS, and Agriculture from the UltraDomain benchmark \cite{qian2024memorag}, and Neurology and Pathology from the MIRAGE benchmark \cite{xiong2024benchmarking}. UltraDomain covers typical RAG applications across different domains, while MIRAGE focuses on medical question answering and domain-specific knowledge coverage. The statistical information is given in the Appendix.

Based on domain consistency and semantic correlation within the texts, we categorize the datasets into three types to enable a comprehensive analysis of the model's adaptability: 
\textbf{\textit{Cross-domain Sparse}} (Mix): Fragmented passages from unrelated domains with weak semantic coherence.
\textbf{\textit{Intra-domain Sparse}} (CS, Agriculture): Domain-specific documents with weak inter-passage context.
\textbf{\textit{Intra-domain Dense}} (Neurology, Pathology): Highly structured medical textbooks with strong semantic continuity from MIRAGE. 
Additionally, we follow the data processing and query procedure of LightRAG, utilizing GPT-4o to generate complex, document-related queries.

\subsubsection{Baselines}
We compared our approach with the state-of-the-art and popular RAG methods. Including text-base RAG: NaiveRAG \cite{gao2023retrieval}, graph-enhanced RAG approaches: GraphRAG \cite{edge2024local}, LightRAG \cite{guo2024lightrag}, HiRAG \cite{huang2025retrieval}, hypergraph-enhanced methods: {Hyper-RAG} \cite{feng2025hyper}. The baseline details are provided in the Appendix.


\subsubsection{Implementation Details}
To ensure fairness and consistency for both the baseline and proposed methods, we validate on five different LLMs for information extraction, question answering, including GPT-4o-mini \cite{achiam2023gpt}, Qwen-Plus \cite{yang2025qwen3}, GLM-4-Air \cite{glm2024chatglm}, DeepSeek-V3 \cite{liu2024deepseek}, and LLaMa-3.3-70B \cite{dubey2024llama}. The result evaluation is default on GPT-4o-mini, as well as the text-embedding-3-small embedding model for vector encoding and retrieval tasks. Unless otherwise specified, all reported results are based on GPT-4o-mini.


\subsubsection{Evaluation Metrics}
Following the recent works, we adopt two evaluation strategies: Selection-based \cite{guo2024lightrag, huang2025retrieval} and Score-based \cite{wang2024leave, feng2025hyper}, providing both relative and absolute perspectives on model performance.
\textbf{\textit{The Selection-based}} evaluation uses LLMs to reports win rates of answer quality between two methods. 
\textbf{\textit{The Score-based}} evaluation employs LLMs to score responses for different methods. Both strategies assess models from six dimensions: Comprehensiveness, Empowerment, Relevance, Consistency, Clarity, and Logical. We report both per-dimension and overall average scores. Detailed evaluation descriptions are in the Appendix.

\begin{table*}[h!]
    \centering
    \scriptsize
    \setlength{\tabcolsep}{1mm}
    \small
    \begin{tabular}{lccccccccccccccccccc}
        \toprule
        & \multicolumn{2}{c}{\textbf{Mix}} & \multicolumn{2}{c}{\textbf{CS}} & \multicolumn{2}{c}{\textbf{Agriculture}} & \multicolumn{2}{c}{\textbf{Neurology}} & \multicolumn{2}{c}{\textbf{Pathology}} \\
        \midrule
        & NaiveRAG & \textbf{Cog-RAG} & NaiveRAG & \textbf{Cog-RAG} & NaiveRAG & \textbf{Cog-RAG} & NaiveRAG & \textbf{Cog-RAG} & NaiveRAG & \textbf{Cog-RAG}\\
        \cmidrule(lr){2-3} \cmidrule(lr){4-5} \cmidrule(lr){6-7} \cmidrule(lr){8-9}\cmidrule(lr){10-11} 
        Comp. & 12.0\% &\textbf{88.0\%} & 4.0\% &\textbf{96.0\%} & 1.0\% &\textbf{99.0\%} & 3.0\% &\textbf{97.0\%} & 6.0\% &\textbf{94.0\%} \\
        Empo. & 10.0\% &\textbf{90.0\%} & 3.0\% &\textbf{97.0\%} & 2.0\% &\textbf{98.0\%} & 1.0\% &\textbf{99.0\%} & 4.0\% &\textbf{96.0\%} \\
        Rele. & 27.0\% &\textbf{73.0\%} & 18.0\% &\textbf{82.0\%} & 6.0\% &\textbf{94.0\%} & 11.0\% &\textbf{89.0\%} & 8.0\% &\textbf{92.0\%} \\
        Cons. & 10.0\% &\textbf{90.0\%} & 4.0\% &\textbf{96.0\%} & 1.0\% &\textbf{99.0\%} & 2.0\% &\textbf{98.0\%} & 4.0\% &\textbf{96.0\%} \\
        Clar. & 23.0\% &\textbf{77.0\%} & 11.0\% &\textbf{89.0\%} & 6.0\% &\textbf{94.0\%} & 6.0\% &\textbf{94.0\%} & 8.0\% &\textbf{92.0\%} \\
        Logi. & 11.0\% &\textbf{89.0\%} & 5.0\% &\textbf{95.0\%} & 1.0\% &\textbf{99.0\%} & 1.0\% &\textbf{99.0\%} & 5.0\% &\textbf{95.0\%} \\
        Overall & 15.5\% &\textbf{84.5\%} & 7.5\% &\textbf{92.5\%} & 2.8\% &\textbf{97.2\%} & 3.2\% &\textbf{96.0\%} & 5.8\% &\textbf{94.2\%} \\
        \midrule
        & GraphRAG & \textbf{Cog-RAG} & GraphRAG & \textbf{Cog-RAG} & GraphRAG & \textbf{Cog-RAG} & GraphRAG & \textbf{Cog-RAG} & GraphRAG& \textbf{Cog-RAG}\\
        \cmidrule(lr){2-3} \cmidrule(lr){4-5} \cmidrule(lr){6-7} \cmidrule(lr){8-9} \cmidrule(lr){10-11} 
        Comp. & 40.0\% &\textbf{60.0\%} & 36.0\% &\textbf{64.0\%} & 32.0\% &\textbf{68.0\%} & 34.0\% &\textbf{66.0\%} & 32.0\% &\textbf{68.0\%} \\
        Empo. & 36.0\% &\textbf{64.0\%} & 35.0\% &\textbf{65.0\%} & 26.0\% &\textbf{74.0\%} & 27.0\% &\textbf{73.0\%} & 23.0\% &\textbf{77.0\%} \\
        Rele. & 45.0\% &\textbf{55.0\%} & 39.0\% &\textbf{61.0\%} & 35.0\% &\textbf{65.0\%} & 37.0\% &\textbf{63.0\%} & 31.0\% &\textbf{69.0\%} \\
        Cons. & 40.0\% &\textbf{60.0\%} & 35.0\% &\textbf{65.0\%} & 29.0\% &\textbf{71.0\%} & 31.0\% &\textbf{69.0\%} & 31.0\% &\textbf{69.0\%} \\
        Clar. & 46.0\% &\textbf{54.0\%} & 36.0\% &\textbf{64.0\%} & 38.0\% &\textbf{62.0\%} & 36.0\% &\textbf{64.0\%} & 30.0\% &\textbf{70.0\%} \\
        Logi. & 39.0\% &\textbf{61.0\%} & 37.0\% &\textbf{63.0\%} & 27.0\% &\textbf{73.0\%} & 33.0\% &\textbf{67.0\%} & 29.0\% &\textbf{71.0\%} \\
        Overall & 41.0\% &\textbf{59.0\%} & 36.3\% &\textbf{63.7\%} & 31.2\% &\textbf{68.8\%} & 33.0\% &\textbf{67.0\%} & 29.5\% &\textbf{70.5\%} \\
        \midrule
        & LightRAG & \textbf{Cog-RAG} & LightRAG & \textbf{Cog-RAG} & LightRAG & \textbf{Cog-RAG} & LightRAG & \textbf{Cog-RAG} & LightRAG & \textbf{Cog-RAG}\\
        \cmidrule(lr){2-3} \cmidrule(lr){4-5} \cmidrule(lr){6-7} \cmidrule(lr){8-9} \cmidrule(lr){10-11} 
        Comp. & 38.0\% &\textbf{62.0\%} & 30.0\% &\textbf{70.0\%} & 23.0\% &\textbf{77.0\%} & 28.0\% &\textbf{72.0\%} & 30.0\% &\textbf{70.0\%} \\
        Empo. & 30.0\% &\textbf{70.0\%} & 26.0\% &\textbf{74.0\%} & 20.0\% &\textbf{80.0\%} & 22.0\% &\textbf{78.0\%} & 25.0\% &\textbf{75.0\%} \\
        Rele. & 36.0\% &\textbf{64.0\%} & 27.0\% &\textbf{73.0\%} & 25.0\% &\textbf{75.0\%} & 28.0\% &\textbf{72.0\%} & 32.0\% &\textbf{68.0\%} \\
        Cons. & 34.0\% &\textbf{66.0\%} & 29.0\% &\textbf{71.0\%} & 21.0\% &\textbf{79.0\%} & 25.0\% &\textbf{75.0\%} & 27.0\% &\textbf{73.0\%} \\
        Clar. & 38.0\% &\textbf{62.0\%} & 24.0\% &\textbf{76.0\%} & 22.0\% &\textbf{78.0\%} & 26.0\% &\textbf{74.0\%} & 26.0\% &\textbf{74.0\%} \\
        Logi. & 35.0\% &\textbf{65.0\%} & 29.0\% &\textbf{71.0\%} & 23.0\% &\textbf{77.0\%} & 26.0\% &\textbf{74.0\%} & 26.0\% &\textbf{74.0\%} \\
        Overall & 35.2\% &\textbf{64.8\%} & 27.5\% &\textbf{72.5\%} & 22.3\% &\textbf{77.7\%} & 25.8\% &\textbf{74.2\%} & 27.7\% &\textbf{72.3\%} \\
        \midrule
        & HiRAG & \textbf{Cog-RAG} & HiRAG & \textbf{Cog-RAG} & HiRAG & \textbf{Cog-RAG} & HiRAG & \textbf{Cog-RAG} & HiRAG & \textbf{Cog-RAG}\\
        \cmidrule(lr){2-3} \cmidrule(lr){4-5} \cmidrule(lr){6-7} \cmidrule(lr){8-9}\cmidrule(lr){10-11} 
        Comp. & 44.0\% &\textbf{56.0\%} & 40.0\% &\textbf{60.0\%} & 41.0\% &\textbf{59.0\%} & 35.0\% &\textbf{65.0\%} & 40.0\% &\textbf{60.0\%} \\
        Empo. & 39.0\% &\textbf{61.0\%} & 36.0\% &\textbf{64.0\%} & 36.0\% &\textbf{64.0\%} & 31.0\% &\textbf{69.0\%} & 37.0\% &\textbf{63.0\%} \\
        Rele. & 45.0\% &\textbf{55.0\%} & 47.0\% &\textbf{53.0\%} & 44.0\% &\textbf{56.0\%} & 35.0\% &\textbf{65.0\%} & 41.0\% &\textbf{59.0\%} \\
        Cons. & 39.0\% &\textbf{61.0\%} & 40.0\% &\textbf{60.0\%} & 37.0\% &\textbf{63.0\%} & 32.0\% &\textbf{68.0\%} & 37.0\% &\textbf{63.0\%} \\
        Clar. & 45.0\% &\textbf{54.0\%} & 50.0\% &\textbf{50.0\%} & 44.0\% &\textbf{56.0\%} & 31.0\% &\textbf{69.0\%} & 40.0\% &\textbf{60.0\%} \\
        Logi. & 40.0\% &\textbf{60.0\%} & 40.0\% &\textbf{60.0\%} & 38.0\% &\textbf{62.0\%} & 31.0\% &\textbf{69.0\%} & 36.0\% &\textbf{64.0\%} \\
        Overall & 42.0\% &\textbf{58.0\%} & 42.2\% &\textbf{57.8\%} & 40.0\% &\textbf{60.0\%} & 32.5\% &\textbf{67.5\%} & 38.5\% &\textbf{61.5\%} \\
        \midrule
        & Hyper-RAG & \textbf{Cog-RAG} & Hyper-RAG & \textbf{Cog-RAG} & Hyper-RAG & \textbf{Cog-RAG} & Hyper-RAG & \textbf{Cog-RAG} & Hyper-RAG & \textbf{Cog-RAG}\\
        \cmidrule(lr){2-3} \cmidrule(lr){4-5} \cmidrule(lr){6-7} \cmidrule(lr){8-9} \cmidrule(lr){10-11} 
        Comp. & 43.0\% &\textbf{57.0\%} & 45.0\% &\textbf{55.0\%} & 49.0\% &\textbf{51.0\%} & 40.0\% &\textbf{60.0\%} & 42.0\% &\textbf{58.0\%} \\
        Empo. & 42.0\% &\textbf{58.0\%} & 43.0\% &\textbf{57.0\%} & 40.0\% &\textbf{60.0\%} & 37.0\% &\textbf{63.0\%} & 37.0\% &\textbf{63.0\%} \\
        Rele. &\textbf{53.0\%} & 47.0\% & 47.0\% &\textbf{53.0\%} & 45.0\% &\textbf{55.0\%} & 46.0\% &\textbf{54.0\%} & 37.0\% &\textbf{63.0\%} \\
        Cons. & 43.0\% &\textbf{57.0\%} & 44.0\% &\textbf{56.0\%} & 44.0\% &\textbf{56.0\%} & 38.0\% &\textbf{62.0\%} & 36.0\% &\textbf{64.0\%} \\
        Clar. &\textbf{56.0}\% & 44.0\% & 48.0\% &\textbf{52.0\%} & 42.0\% &\textbf{58.0\%} & 41.0\% &\textbf{59.0\%} & 32.0\% &\textbf{68.0\%} \\
        Logi. & 44.0\% &\textbf{56.0\%} & 46.0\% &\textbf{54.0\%} & 43.0\% &\textbf{57.0\%} & 35.0\% &\textbf{65.0\%} & 37.0\% &\textbf{63.0\%} \\
        Overall & 46.8\% &\textbf{53.2\%} & 45.5\% &\textbf{54.5\%} & 43.8\% &\textbf{56.2\%} & 39.5\% &\textbf{60.5\%} & 36.8\% &\textbf{63.2\%} \\
        
        \bottomrule
    \end{tabular}
    \caption{Average win rates of six evaluation metrics across five datasets. The comparison is made between baselines and Cog-RAG. Among them, we refer to the six metrics as Comp. (Comprehensiveness), Empo. (Empowerment), Rele. (Relevance), Cons. (Consistency), Clar. (Clarity), and Logi. (Logical).}

    \label{tab1}
\end{table*}

\subsection{Main Results}
Our primary results are presented in Table \ref{tab1} and Figure \ref{fig3}, and more results are provided in the Appendix. Cog-RAG consistently outperforms all baselines across multiple dimensions. Additionally, we have several key insights:

1) \textbf{Knowledge graphs can enhance RAG to model a broader scope of information.} 
Graph-enhanced methods, represented by GraphRAG and LightRAG, demonstrate significant advantages over the conventional NaiveRAG, primarily due to the modeling of graph structures. In contrast, NaiveRAG relies solely on vector similarity and fails to account for these structured semantic relations. Hypergraph-enhanced approaches, such as Hyper-RAG and Cog-RAG, offer a more comprehensive modeling of knowledge structures that extend beyond pairwise relations, demonstrating superior potential in knowledge representation.

2) \textbf{Cog-RAG outperforms the baselines across all kinds of evaluation datasets and LLMs.} 
For \textit{Selection-based results} in Table \ref{tab1}, we can see that in cross-domain sparse settings, 
both Hyper-RAG and Cog-RAG utilize hypergraphs to capture high-order relations, resulting in an average improvement of over 10.0\% compared to graph-based methods.
In intra-domain sparse datasets, Cog-RAG outperforms HiRAG by 15.6\% and 20.0\%, benefiting from multi-hyperedge propagation that uncovers latent themes and entity relations.
In intra-domain dense medical corpora, Cog-RAG achieves the most significant gains. Through dual hypergraph modeling and cognitive-inspired retrieval, 
enhancing the alignment and aggregation of theme and fine-grained details. Compared to Hyper-RAG, it improves by 21.0\% and 26.4\%, respectively. 
For \textit{Score-based results}, Figure \ref{fig3} objectively presents the evaluation results across six dimensions on five LLMs. The results demonstrate that Cog-RAG achieves consistent and significant improvements over baseline methods in all dimensions. Moreover, when applying different LLMs for indexing and answering, it still exhibits clear advantages, highlighting its structural effectiveness.

3) \textbf{Dual-hypergraph alignment enhances knowledge representation and semantic consistency.}
Inspired by human top-down cognitive pathways, Cog-RAG utilizes a dual hypergraph structure to align macro to micro knowledge. 
Specifically, for \textit{Selection-based results} in Intra-domain Dense scenario, Cog-RAG improves by 35.0\% and 23.0\% compared to the entity-level hierarchical method HiRAG. For \textit{Score-based results}, Cog-RAG outperforms HiRAG and Hyper-RAG by 1.37 and 1.20 on neurology datasets, significantly enhancing the model's ability to handle knowledge-intensive domains and ensuring semantic consistency. 

\subsection{Ablation Study}
This section conducts ablation studies to evaluate the contribution of each core component in Cog-RAG: the theme, entity hypergraph, and two-stage retrieval strategy. Table \ref{tab2} shows the results by Scoring-based evaluation, and Selection-based results can be found in the Appendix. The results from three types of representative datasets are summarized below.

1) \textbf{Effectiveness of the Entity Hypergraph.}
Removing the entity hypergraph leads to a significant decrease in performance on all three types of datasets, especially on the Mix dataset. This indicates its critical role in capturing fine-grained semantic relations within chunks. This effect is consistently observed across domains, confirming that intra-chunk entity-level modeling can enhance the representation of local knowledge.

2) \textbf{Effectiveness of the Theme Hypergraph.}
Excluding the theme hypergraph causes a moderate decrease (drop 1.19 on CS and 1.14 on Neurology), highlighting its role in modeling global theme structures across chunks. The benefit is particularly noticeable in intra-domain tasks, where maintaining coherent theme alignment helps with cross-chunk reasoning and retrieval.
However, on the Mix dataset, using only the theme hypergraph leads to performance degradation (from 85.39 to 76.58), indicating that in cross-domain sparse and weakly structured scenarios, theme relations may introduce noise that interferes with retrieval and answering.

3) \textbf{Effectiveness of the Two-Stage Retrieval.}
Bypassing this component (by directly concatenating information from both the theme and entity hypergraphs and inputting it into LLMs) leads to consistent performance drops. This highlights the importance of the two-stage retrieval, especially in knowledge-intensive scenarios where global semantic guidance followed by entity-level refinement enables more accurate and coherent retrieval.

\begin{table}[t!]
    \centering
    \small
    \begin{tabular}{lccc}
    \toprule
    \multirow{2}{*}{\textbf{Models}} & \textbf{Mix} & \textbf{CS} & \textbf{Neurology} \\
    \cmidrule(r){2-2} \cmidrule(r){3-3} \cmidrule(r){4-4}
    & (Overall) & (Overall) & (Overall) \\
    \midrule
    \textsc{Cog-RAG} & 85.39 & 87.07 & 86.55 \\
    \textit{w./o. Entity Hypergraph} & 76.58 & 84.58 & 84.49 \\
    \textit{w./o. Theme Hypergraph} & 84.82 & 85.88 & 85.41 \\
    \textit{w./o. Two-Stage Retrieval} & 84.88 & 86.41 & 86.18 \\
    \bottomrule
    \end{tabular}
    \caption{Ablation study on different datasets by scoring, where w./o. indicates without the part of the method.}
    \label{tab2}
\end{table}


\subsection{Hypergraph Visualization}
\begin{figure}[t]
    \centering\includegraphics[width = \linewidth]{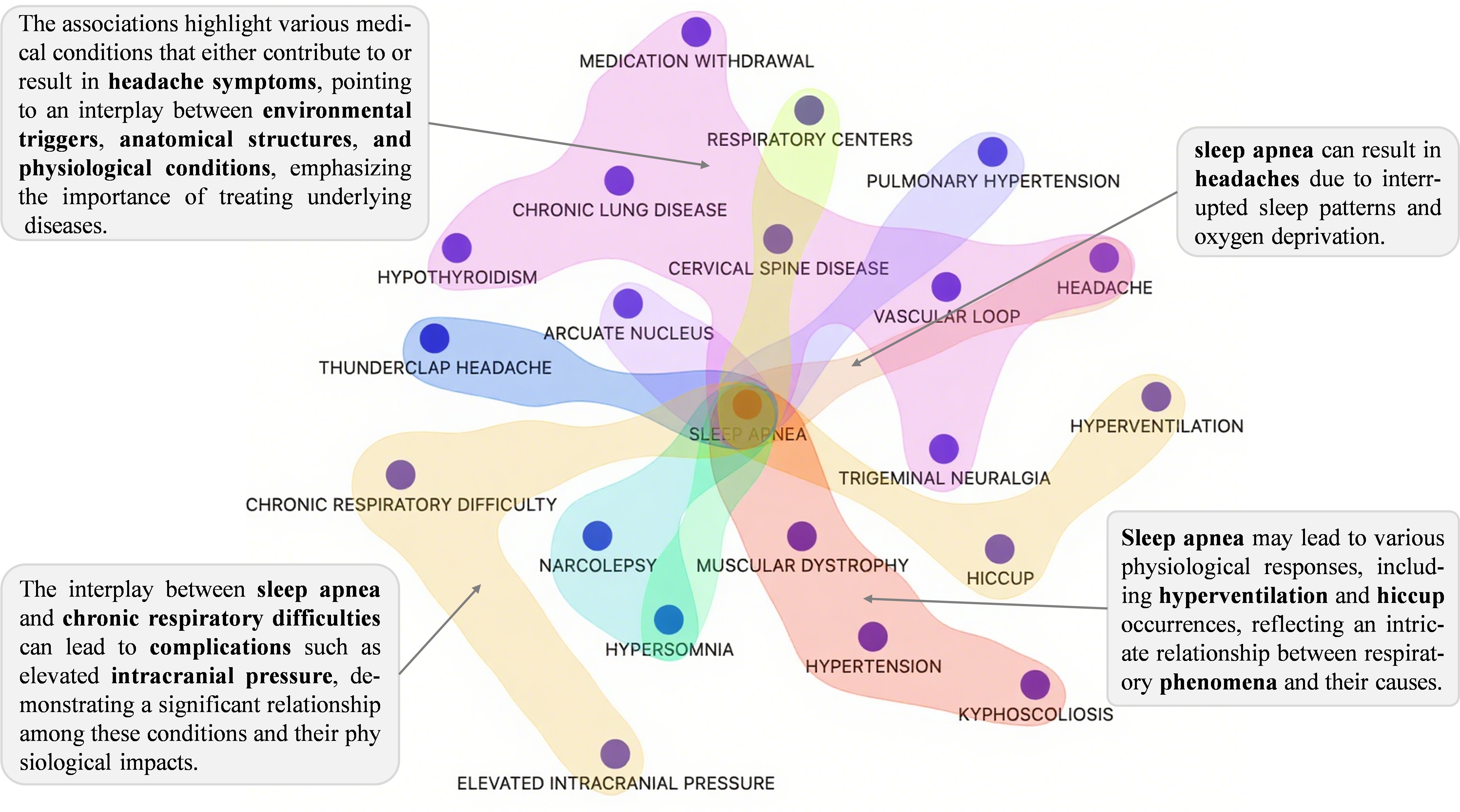}
    \caption{Entity Hypergraph Visualization.}
    \label{fig4}
\end{figure}

In the Neurology dataset, Figure \ref{fig4} visualized the relations of Sleep Apnea in the entity hypergraph. It illustrates the complex relations between Sleep Apnea and multiple related entities such as Chronic Lung Disease, Headache, and Respiratory Centers. It captures not only pairwise relations but also reveals multi-entity dependencies beyond pairs.
As observed, the complex hypergraph among Hypertension, Sleep Apnea, Kyphoscoliosis, and Muscular Dystrophy illustrates various health risks and respiratory challenges connected to sleep quality and disorders, affecting overall wellness.



\section{Why is Cog-RAG Effective?}
\subsection{Theme-Aligned vs. Graph / Hypergraph Index}
Graph and hypergraph-enhanced RAG mainly focus on modeling local entity-level relations within document chunks, making them less effective for tasks that require global semantic reasoning. In contrast, Cog-RAG introduces a dual-hypergraph structure that supports alignment from global themes to fine-grained entities, leading to improved contextual grounding and response consistency. 
Notably, our analysis reveals that the theme hypergraph is particularly beneficial in structured, domain-specific settings, while it may introduce noise in loosely structured, open-domain scenarios. This suggests further opportunities for dynamic filtering and graph construction.

\subsection{Cognitive-Inspired vs. Conventional Retrieval}
Conventional RAG systems rely on single-stage retrieval, which merges all retrieved content into LLMs. This design often leads to incomplete or noisy evidence aggregation for complex knowledge-intensive tasks. The cognitive-inspired two-stage retrieval strategy
enables top-down semantic alignment and aggregation, providing more accurate knowledge support and reducing redundant information.

\section{Conclusion}
Inspired by human cognitive pathways, this paper introduces Cog-RAG, which enhances LLM responses by integrating dual-hypergraph structures and a cognitive-inspired two-stage retrieval mechanism.
Cog-RAG enables hierarchical knowledge modeling and semantic alignment at both macro-thematic and micro-entity levels, addressing issues of information loss and semantic gaps inherent in graph-based methods.
Experimental results show that Cog-RAG significantly outperforms state-of-the-art methods across various types of datasets on knowledge-intensive tasks.

\section{Acknowledgement}
This work was supported by the National Natural Science Foundation of China under Grant Nos. 62088102 and U24A20252, the Key Research and Development Program of Shaanxi Province of China under Grant Nos. 2024PT-ZCK-66 and 2024CY2-GJHX-48.

\bibliography{mybib}

\clearpage
\section{APPENDIX}
\vspace{3em}

\section{A. Experimental Datasets}
\begin{table}[h]
    \centering
    \resizebox{\linewidth}{!}{
    \begin{tabular}{lccccc}
    \toprule
    \textbf{Statistics} & \textbf{Mix} & \textbf{CS} & \textbf{Agriculture}  & \textbf{Neurology} & \textbf{Pathology} \\
    \midrule
    Total Documents & 61 & 10 & 12 & 1 & 1 \\
    Total Chunks &  560 & 1992 & 1813 & 1790 & 824 \\
    Total Tokens & 615,355 & 2,190,803 & 1,993,515 & 1,968,716 & 905,760 \\
    \bottomrule
    \end{tabular}}
    \caption{Statistical information of the datasets}
    \label{tab3}
\end{table}

Table 1 presents the statistical information of the five datasets.

\section{B. Details of Baselines}
We compared our approach with the state-of-the-art and popular RAG methods. The specific description of the method is as follows:

\textbf{NaiveRAG}: The baseline of the standard RAG systems, which segments texts into chunks and stores them as embeddings in a vector database. During retrieval, relevant chunks are directly matched via vector similarity.

\textbf{GraphRAG}: A standard graph-enhanced RAG method that employs LLMs to extract entities and relations from texts as nodes and edges of the graph. Entities are further clustered to generate community reports, which are traversed during retrieval to obtain global information.

\textbf{LightRAG}: A graph-enhanced RAG method that integrates graph structures with vector-based representations. It employs a dual-level retrieval strategy to retrieve information from both nodes and edges within the graph knowledge.

\textbf{HiRAG}: A graph-enhanced RAG approach that builds a hierarchical graph via multi-level semantic clustering, enabling hierarchical indexing and retrieval of text knowledge.

\textbf{Hyper-RAG}: A standard Hypergraph-enhanced RAG method that uses hyperedges to represent both paired low-order relations and beyond-paired high-order relations.

\begin{figure*}[t!]
    \centering\includegraphics[width = 0.57\linewidth]{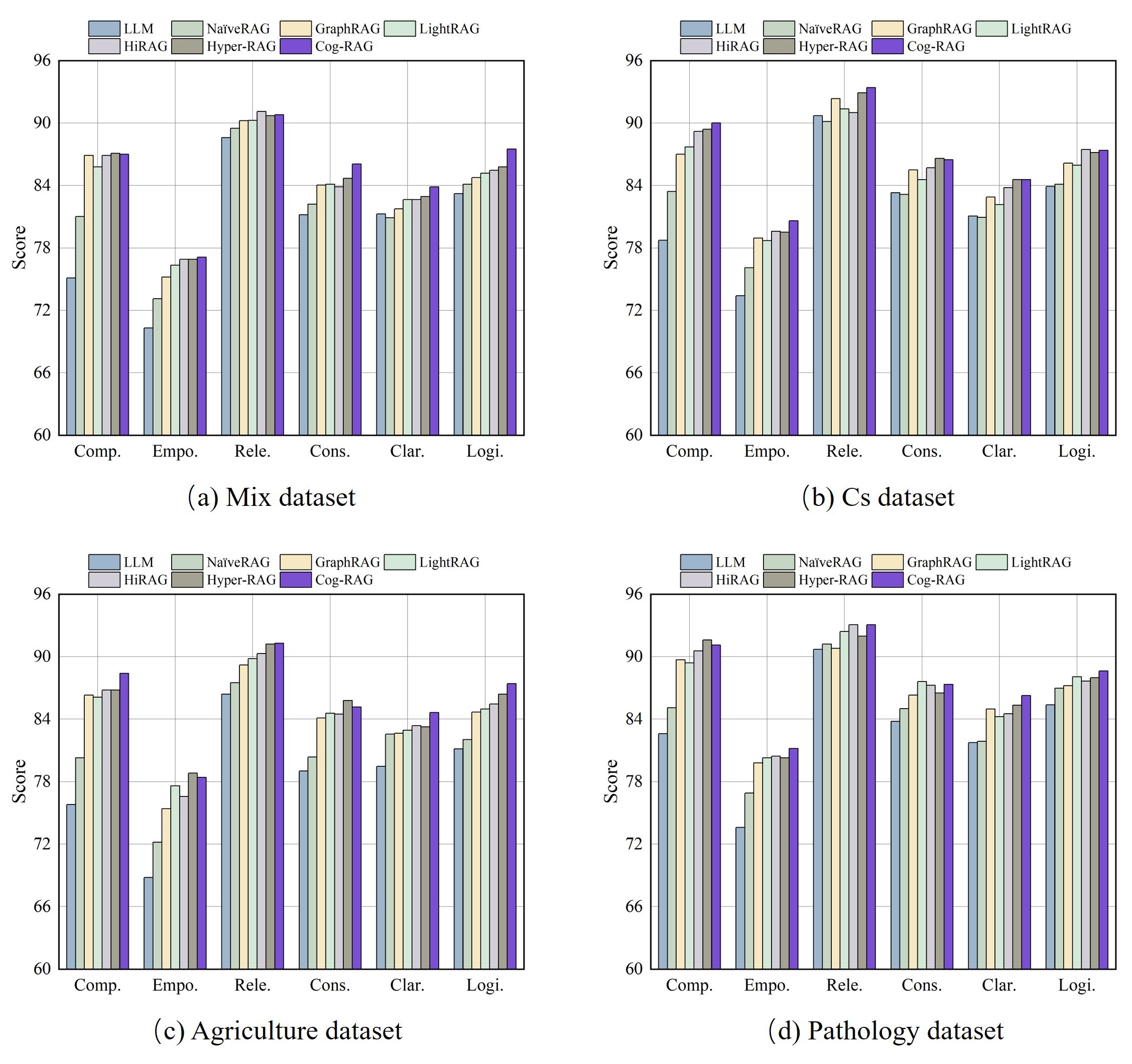}
    \caption{Comparison results on different metrics.}
    \label{app_fig1}
\end{figure*}

\begin{table*}[t!]
    \centering
    \small
    \begin{tabular}{lccccccccc}
        \toprule
        & \multicolumn{2}{c}{\textbf{CS}} & \multicolumn{2}{c}{\textbf{Mix}} & \multicolumn{2}{c}{\textbf{Neurology}} \\ 
        \midrule
        &  \makecell{w./o. Entity \\ Hypergraph} & \textbf{Cog-RAG} & \makecell{w./o. Entity \\ Hypergraph} & \textbf{Cog-RAG} & \makecell{w./o. Entity \\ Hypergraph} & \textbf{Cog-RAG}\\
        \cmidrule(lr){2-3} \cmidrule(lr){4-5} \cmidrule(lr){6-7} 
        Comp. & 24.0\% & \textbf{76.0\%} & 34.0\% & \textbf{66.0\%} & 30.0\% & \textbf{70.0\%} \\
        Empo. & 20.0\% & \textbf{80.0\%} & 41.0\% & \textbf{59.0\%} & 23.0\% & \textbf{77.0\%} \\
        Rele. & 29.0\% & \textbf{71.0\%} & 31.0\% & \textbf{69.0\%} & 30.0\% & \textbf{70.0\%} \\
        Cons. & 24.0\% & \textbf{76.0\%} & 41.0\% & \textbf{59.0\%} & 26.0\% & \textbf{74.0\%} \\
        Clar. & 25.0\% & \textbf{75.0\%} & 25.0\% & \textbf{75.0\%} & 30.0\% & \textbf{70.0\%} \\
        Logi. & 24.0\% & \textbf{76.0\%} & 44.0\% & \textbf{56.0\%} & 28.0\% & \textbf{72.0\%} \\
        Overall & 24.0\% & \textbf{76.0\%} & 36.0\% & \textbf{64.0\%} & 28.0\% & \textbf{72.0\%} \\
        \midrule
        & \makecell{w./o. Theme \\ Hypergraph} & \textbf{Cog-RAG} & \makecell{w./o. Theme \\ Hypergraph} & \textbf{Cog-RAG} & \makecell{w./o. Theme \\ Hypergraph} & \textbf{Cog-RAG}\\
        \cmidrule(lr){2-3} \cmidrule(lr){4-5} \cmidrule(lr){6-7} 
        Comp. & 46.0\% & \textbf{54.0\%} & 41.0\% & \textbf{59.0\%} & 43.0\% & \textbf{57.0\%} \\
        Empo. & 45.0\% & \textbf{55.0\%} & 39.0\% & \textbf{61.0\%} & 39.0\% & \textbf{61.0\%} \\
        Rele. & 47.0\% & \textbf{53.0\%} & 39.0\% & \textbf{61.0\%} & 42.0\% & \textbf{58.0\%} \\
        Cons. & 45.0\% & \textbf{55.0\%} & 41.0\% & \textbf{59.0\%} & 41.0\% & \textbf{59.0\%} \\
        Clar. & \textbf{57.0\%} & 43.0\% & 39.0\% & \textbf{61.0\%} & 41.0\% & \textbf{59.0\%} \\
        Logi. & 48.0\% & \textbf{52.0\%} & 40.0\% & \textbf{60.0\%} & 41.0\% & \textbf{59.0\%} \\
        Overall & 48.0\% & \textbf{52.0\%} & 40.0\% & \textbf{60.0\%} & 41.0\% & \textbf{59.0\%} \\
        \midrule
        & \makecell{w./o. Two-Stage\\ Retrieval} & \textbf{Cog-RAG} & \makecell{w./o. Two-Stage\\ Retrieval} & \textbf{Cog-RAG} & \makecell{w./o. Two-Stage\\ Retrieval} & \textbf{Cog-RAG}\\
        \cmidrule(lr){2-3} \cmidrule(lr){4-5} \cmidrule(lr){6-7} 
        Comp. & 44.0\% & \textbf{56.0\%} & 46.0\% & \textbf{54.0\%} & 43.0\% & \textbf{57.0\%} \\
        Empo & 40.0\% & \textbf{60.0\%} & 44.0\% & \textbf{56.0\%} & 40.0\% & \textbf{60.0\%} \\
        Rele. & \textbf{51.0\%} & 49.0\% & 45.0\% & \textbf{55.0\%} & 39.0\% & \textbf{61.0\%} \\
        Cons. & 39.0\% & \textbf{61.0\%} & 47.0\% & \textbf{53.0\%} & 44.0\% & \textbf{56.0\%} \\
        Clar. & 50.0\% & \textbf{50.0\%} & 49.0\% & \textbf{51.0\%} & 49.0\% & \textbf{51.0\%} \\
        Logi. & 38.0\% & \textbf{62.0\%} & 47.0\% & \textbf{53.0\%} & 45.0\% & \textbf{55.0\%} \\
        Overall & 44.0\% & \textbf{56.0\%} & 46.0\% & \textbf{54.0\%} & 43.0\% & \textbf{57.0\%} \\        
        \bottomrule
    \end{tabular}
    \caption{Average win rates on three datasets. The comparison is made between ablation and Cog-RAG. }
    \label{tab5}
\end{table*}

\section{C. Details of Evaluation Metrics}
We evaluated the proposed method and baseline from six dimensions, including Comprehensiveness, Empowerment, Relevance, Consistency, Clarity, and Logical coherence. 

For Selection-based evaluation, we report win rates to conduct a qualitative comparison. We also alternated the answer order of each pair of methods in the prompts and calculated the average result for a fair comparison. The metrics details are described below:

\textbf{Comprehensiveness}: How much detail does the answer provide to cover all aspects and details of the question?
\textbf{Empowerment}: How well does the answer help the reader understand and make informed judgments about the topic?
\textbf{Relevance}: How precisely does the answer address the core aspects of the question without including unnecessary information?
\textbf{Consistency}: How well does the system integrate and synthesize information from multiple sources into a logically flowing response?
\textbf{Clarity}: How well does the system provide complete information while avoiding unnecessary verbosity and redundancy?
\textbf{Logical}: How well does the system maintain consistent logical arguments without contradicting itself across the response?

For Score-based evaluation, we use LLMs to score responses for quantitative assessment, with the following specific indicators:

\textbf{Comprehensiveness (0-100)}: Measure whether the answer comprehensively covers all key aspects of the question and whether there are omissions.
\textbf{Empowerment (0-100)}: Measure the credibility of the answer and whether it convinces the reader that it is correct. High confidence answers often cite authoritative sources or provide sufficient evidence.
\textbf{Relevance (0-100)}: Measure whether the content of the answer is closely related to the question, and whether it stays focused on the topic without digression.          
\textbf{Consistency (0-100)}: Measure whether the answer is logically organized, flows smoothly, and whether the parts of the answer are well connected and mutually supportive.
\textbf{Clarity (0-100)}: Measure whether the answer is expressed in a clear, unambiguous, and easily understandable manner, using appropriate language and definitions.
\textbf{Logical (0-100)}: Measure whether the answer is coherent, clear, and easy to understand.

For each evaluation dimension, we define five discrete rating levels, each associated with clear scoring criteria to ensure consistency and transparency. As an illustration, the Comprehensiveness dimension is rated as follows:
\textbf{Level 1 (0-20)}: The answer is extremely one-sided, leaving out key parts or important aspects of the question.
\textbf{Level 2 (20-40)}: The answer has some content but misses many important aspects and is not comprehensive enough.
\textbf{Level 3 (40-60)}: The answer is more comprehensive, covering the main aspects of the question, but there are still some omissions.
\textbf{Level 4 (60-80)}: The answer is comprehensive, covering most aspects of the question with few omissions.
\textbf{Level 5 (80-100)}: The answer is extremely comprehensive, covering all aspects of the question with no omissions, enabling the reader to gain a complete understanding.

\section{D. Additional Experiment Results}
\subsection{D.1 Comparison Experiment Results}
Figure 1 shows the results of comparison experiments under different evaluation dimensions on the Mix, CS, Agriculture, and Pathology datasets.

\subsection{D.2 Ablation Experiment Results}
Table 2 presents the results of the selection-based evaluation on three types of representative datasets, including Mix, CS, and Neurology. The results also show the effectiveness of each component in Cog-RAG.

\section{E. Retrieval Efficiency Analysis}
We conduct a comparative analysis of retrieval efficiency. Figure 2 illustrates the trade-off between retrieval time and final answer score across different methods. Notably, Cog-RAG achieves the highest overall score while maintaining low retrieval overhead, highlighting its superior balance between performance and efficiency.
\begin{figure}[h!]
    \centering\includegraphics[width = 0.85\linewidth]{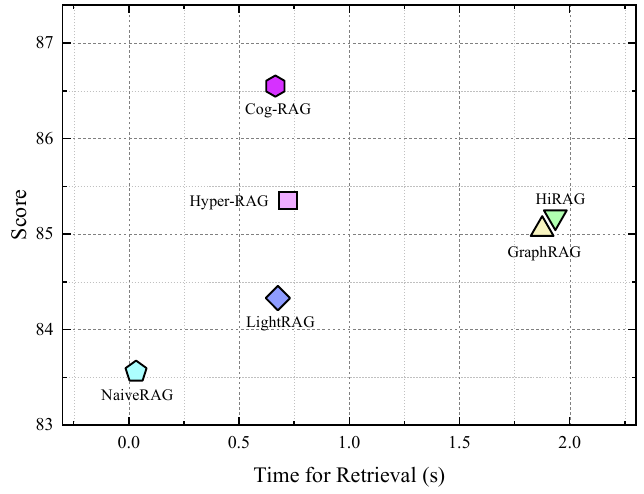}
    \caption{Comparison results on different metrics.}
    \label{app_fig2}
\end{figure}

\section{F. Prompt Templates in Cog-RAG}
\subsection{F.1 Extracting Themes, Key Entities}
\begin{myprompt_single}{Extracting Themes}
    \textbf{Formulation}: $\mathcal{P}_\text{ext\_theme}$\newline 
    
    \textbf{Prompt: }\textit{
    Summarize the primary theme of the text document. This summary should capture the essence of the document’s core conflict, main idea, or narrative arc. Ensure that the summary highlights key moments, changes, or shifts in the document, extract the following information:\newline
    - theme\_description: A sentence that describes the primary theme of the text document, reflecting the main conflict, resolution, or key message.
}
\end{myprompt_single}

\begin{myprompt_single}{Extracting Key Entities}
    \textbf{Formulation}: $\mathcal{P}_\text{ext\_key}$\newline 
    
    \textbf{Prompt: }\textit{
    From the theme identified in $\mathcal{P}_\text{ext\_key}$, identify all key entities in each text document. \newline
    For each identified key entity, extract the following information:\newline
    - key\_entity\_name: Name of the key entity, use same language as input text.  If English, capitalized the name.\newline
    - key\_entity\_type: Type of the key entity, such as person, concept, object, event, emotion, symbol.\newline
    - key\_entity\_description: Comprehensive description of the entity's attributes and activities.\newline
    - key\_score: A score from 0 to 100 indicating the importance of the entity in the text.
}
\end{myprompt_single}

\subsection{F.2 Extracting Entities, Relations and Keywords}
\begin{myprompt_single}{Extracting Entities}
    \textbf{Formulation}: $\mathcal{P}_\text{ext\_entity}$\newline \newline
    \textbf{Prompt: }\textit{
    Identify all entities. For each identified entity, extract the following information:\newline
    - entity\_name: Name of the entity, use same language as input text. If English, capitalized the name.\newline
    - entity\_type: One of the following types: organization, person, geo, event, role, concept.\newline
    - entity\_description: Comprehensive description of the entity's attributes and activities.\newline
    - additional\_properties: Other attributes possibly associated with the entity, like time, space, emotion, motivation, etc.
}
\end{myprompt_single}

\begin{myprompt_single}{Extracting Low-order Relations}
    \textbf{Formulation}: $\mathcal{P}_\text{ext\_low}$\newline \newline
    \textbf{Prompt: }\textit{
    From the entities identified in $\mathcal{P}_\text{ext\_entity}$, identify all pairs of (source\_entity, target\_entity) that are *clearly related* to each other.\newline
    For each pair of related entities, extract the following information:\newline
    - entities\_pair: The name of source entity and target entity, as identified in $\mathcal{P}_\text{ext\_entity}$.\newline
    - low\_order\_relationship\_description: Explanation as to why you think the source entity and the target entity are related to each other.\newline
    - low\_order\_relationship\_keywords: Keywords that summarize the overarching nature of the relationship, focusing on concepts or themes rather than specific details.\newline
    - low\_order\_relationship\_strength: A numerical score indicating the strength of the relationship between the entities.
}
\end{myprompt_single}

\begin{myprompt_single}{Extracting Theme Keywords from Query}
    \textbf{Formulation}: $\mathcal{P}_\text{keywords}$\newline \newline
    \textbf{Prompt: }\textit{
    You are a helpful assistant tasked with identifying theme-level keywords in the user's query.\newline
    ---Goal---\newline
    Given the query, list theme-level keywords. theme-level keywords focus on overarching concepts or themes.
}
\end{myprompt_single}

\begin{myprompt_double}{Extracting High-order Relations}
    \textbf{Formulation}: $\mathcal{P}_\text{ext\_high}$\newline\newline
    \textbf{Prompt: }\textit{
    For the entities identified in $\mathcal{P}_\text{ext\_entity}$, based on the entity pair relationships in $\mathcal{P}_\text{ext\_low}$, find connections or commonalities among multiple entities and construct high-order associated entity set as much as possible.\newline
    Extract the following information from all related entities, entity pairs:\newline
    - entities\_set: The collection of names for elements in high-order associated entity set.\newline
    - high\_order\_relationship\_description: Use the relationships among the entities in the set to create a detailed, smooth, and comprehensive description that covers all entities in the set, without leaving out any relevant information.\newline
    - high\_order\_relationship\_generalization: Summarize the content of the entity set as concisely as possible.\newline
    - high\_order\_relationship\_keywords: Keywords that summarize the overarching nature of the high-order association, focusing on concepts or themes rather than specific details.\newline
    - high\_order\_relationship\_strength: A numerical score indicating the strength of the association among the entities in the set.
    }
\end{myprompt_double}

\subsection{F.3 Theme Alignment to Entity}
\begin{myprompt_double}{Theme Alignment to Entity}
    \textbf{Formulation}: $\mathcal{P}_\text{algin}(q,\mathcal{A}_\text{theme})$\newline\newline
    \textbf{Prompt: }\textit{
    You are a helpful assistant tasked with extracting entity-level keywords from the query $q$ that are theme-aligned.\newline
    ---Goal---\newline
    Given the query and theme, extract specific entities and concrete terms that relate to the identified theme.\newline
    - Extract keywords that are concrete and entity-focused.\newline
    - Prioritize keywords aligned with the theme.\newline
    Here are the question: $q$\newline
    Here are the theme: $\mathcal{A}_\text{theme}$
    }
\end{myprompt_double}

\subsection{F.4 Evaluation Metrics}
\begin{myprompt_double}{Selection-based Evaluation}
    \textbf{Formulation}: $\mathcal{P}_\text{eval\_scoring}(q, \mathcal{A}_1, \mathcal{A}_2)$\newline 
    $q$ denotes user query, $\mathcal{A}_1$ and $\mathcal{A}_2$ denotes the response from two approaches. \newline\newline
    \textbf{Prompt: }\textit{You will evaluate two answers to the same question based on six criteria: Comprehensiveness, Empowerment, Relevance, Consistency, Clarity, and Logical.\newline
    ---Goal---\newline
    You will evaluate two answers to the same question by using the relevant documents based on six criteria: Comprehensiveness, Empowerment, Relevance, Consistency, Clarity, and Logical.\newline
    -Comprehensiveness: How much detail does the answer provide to cover all aspects and details of the question?\newline
    -Empowerment: How well does the answer help the reader understand and make informed judgments about the topic?\newline
    ...,\newline
    -Logical: How well does the system maintain consistent logical arguments without contradicting itself across the response?\newline\newline
    For each criterion, choose the better answer (either Answer 1 or Answer 2) and explain why. Then, select an overall winner based on these six categories.\newline
    Here are the question: $q$\newline
    Here are the two answers: \newline
    Answer 1: $\mathcal{A}_1$;\newline
    Answer 2: $\mathcal{A}_2$\newline
    Evaluate both answers using the six criteria listed above and provide detailed explanations for each criterion.
}
\end{myprompt_double}

\clearpage

\begin{myprompt_double}{Scoring-based Evaluation}
    \textbf{Formulation}: $\mathcal{P}_\text{eval\_scoring}(q, \mathcal{A}, \mathcal{C})$\newline 
    $q$ denotes user query, $\mathcal{A}$ denotes LLM response, $\mathcal{C}$ denotes the original text chunk that generated the question. \newline\newline
    \textbf{Prompt: }\textit{You are an expert tasked with evaluating answers to the questions by using the relevant documents based on five criteria: Comprehensiveness, Diversity, Empowerment, Logical, and Readability.\newline\newline
    ---Goal---\newline
     You will evaluate tht answers to the questions by using the relevant documents based on on six criteria: Comprehensiveness, Empowerment, Relevance, Consistency, Clarity, and Logical.\newline\newline
    -Comprehensiveness-\newline
    Measure whether the answer comprehensively covers all key aspects of the question and whether there are omissions.\newline
    Level   $|$ score range $|$ description\newline
    Level 1 $|$ 0-20   $|$ The answer is extremely one-sided, leaving out key parts or important aspects of the question.\newline
    Level 2 $|$ 20-40  $|$ The answer has some content, but it misses many important aspects of the question and is not comprehensive enough.\newline
    Level 3 $|$ 40-60  $|$ The answer is more comprehensive, covering the main aspects of the question, but there are still some omissions.\newline
    Level 4 $|$ 60-80  $|$ The answer is comprehensive, covering most aspects of the question, with few omissions.\newline
    Level 5 $|$ 80-100 $|$ The answer is extremely comprehensive, covering all aspects of the question with no omissions, enabling the reader to gain a complete understanding.\newline
    ...,\newline\newline
    For each indicator, please give the problem a corresponding Level based on the description of the indicator, and then give a score according to the score range of the level.\newline
    Here are the question: $q$\newline
    Here are the relevant document: $\mathcal{C}$\newline
    Here are the answer: $\mathcal{A}$\newline
    Evaluate all the answers using the six criteria listed above, for each criterion, provide a summary description, give a Level based on the description of the indicator, and then give a score based on the score range of the level.
}
\end{myprompt_double}

\end{document}